\documentclass[english,amsmath,fleqn,amssymb,twocolumn,aps,prl,10pt,superscriptaddress,floatfix,showpacs]{revtex4-1}
\usepackage{hyperref}
\usepackage{booktabs}
\usepackage{balance}
\usepackage{color}
\usepackage{units}

\newcommand{\mathsym}[1]{{}}
\newcommand{\unicode}[1]{{}}
\newcommand{\exv}[1]{{\left\langle #1 \right\rangle}}
\newcommand{\norm}[1]{{\left| #1 \right|}}
\newcommand{\hide}[1]{}

\newcommand{\eq}[1]{Eq.\,(\ref{#1})}

\newcommand{\tab}[1]{Table\,\ref{#1}}
\newcommand{\re}[1]{{{\rm Re}\!\left[ #1 \right]}}

\begin{document}
\title{New Quantum Bounds for Inequalities involving Marginal Expectations}
\author{Elie Wolfe}
\email{wolfe@phys.uconn.edu}
\affiliation{Department of Physics, University of Connecticut, Storrs, CT 06269}

\author{S.F. Yelin}
\affiliation{Department of Physics, University of Connecticut, Storrs, CT 06269}
\affiliation{ITAMP, Harvard-Smithsonian Center for Astrophysics, Cambridge MA 02138}

\date{\today}
\pacs{03.65.Ud,03.65.Ta,03.67.Ac}

\begin{abstract}
We review, correct, and develop an algorithm to determine arbitrary Quantum Bounds based on the seminal work of Tsirelson [Lett. Math. Phys. \textbf{4}, 93 (1980)]. The vast potential of this algorithm is freshly demonstrated by deriving both new number-valued Quantum Bounds, as well as identifying a new class of function-valued Quantum Bounds. Those results facilitate an 8-dimensional Volume Analysis of Quantum Mechanics which extends the work of Cabello [PRA \textbf{72} (2005)]. Finally we contrast  these bounds to those defined by the first-level criteria of Navascu\'{e}s et al [NJP \textbf{10} (2008)], proving our function-valued quantum bounds to be relatively more complete.
\end{abstract}
\maketitle

\section{I. Introduction}

Quantum Mechanics (QM) is an inextricably nonlocal \cite{epr} theory: it is fundamentally incompatible with Local Hidden Variable Models (LHVM). This is perhaps best evidenced by the violation of Bell inequalities \cite{epr,Bell1964,CHSH}, although other manifestations of nonlocality are also known \cite{ks18,pentagrams,cabello_qutrit,mosca_algorithms_revisited}. On the other hand it is now appreciated that QM is less than maximally nonlocal, in the sense that causality-respecting, i.e. no-signaling-faster-than-light (NOSIG) frameworks allow for violation of the Bell inequalities even beyond the Quantum limit \cite{cabello05,pr,scarinilecture}. Many recent works have offered information-theoretic explanations to account for the discrepancy between those statistics a-priori permitted by General Relativity and those a-posteriori limited by quantum theory \cite{vandam,swapping,infocaus}.
%
%
%
%

In this work we investigate the distinctions between LHVM, QM, and NOSIG through gedankenexperiments consisting of parties \((A, B, C\ldots)\), each having access to measurement apparatuses \((A_0, A_1\ldots B_0,B_1\ldots)\) capable of finite measurement outcomes. Here we consider only binary measurement outcomes, which we take to be \(\pm 1\). Practically this amounts to spin measurements on physically separated qubits, where the $A_0$ is a spin measure relative to angle $\theta_{A_0}$ and so forth. The statistical quantities of interest are the marginal probabilities \(P(A_0=-1)\), \(P(A_0=+1)\) for all the measurement apparatuses for all the parties, as well as the joint coincidence counts such as \(P(A_i=B_j)\) which we can equivalently express more generally as \(P\left( (A_i \times B_j \times C_k \times ...) = 1\right)\) for all \(i,j,k,...\) Here we follow the conventional notation of {\em expectation values} such that \(\langle A_i\rangle \equiv P(A_i=1)-P(A_i=-1)\) and \(\langle A_i=B_j\rangle \equiv  \langle A_i\cdot B_j\rangle\). We'll focus on inequalities (bounds) pertaining to linear combinations of the expectation values, such as 
\begin{align}\label{eq:simple}
	\hspace{-23pt}\left\langle A_0\cdot B_0\right\rangle
	+\left\langle A_0\cdot B_1\right\rangle
	+\left\langle A_1\cdot B_0\right\rangle
	-\left\langle A_1\cdot B_1\right\rangle
\,\leq\,\gamma
\end{align}

LHVM posits that all measurements merely reveal pre-established values. One can demonstrate \cite{scarinilecture,roberts_thesis} that all statistics consistent with LHVM ``fit" within \eq{eq:simple} upper-bounded by \(\gamma={2}\). One can also prove \cite{pr,scarinilecture} that NOSIG, in contrast, permits \eq{eq:simple} to go all the way up to \(\gamma={4}\). The quantum bound for \eq{eq:simple} has been proven  \cite{tsirelson80} to be exactly \(\gamma={2\sqrt{2}}\), in between the bounds of LHVM and NOSIG. We'll refer to the minimal binary and dichotomic scenario implicitly referenced in \eq{eq:simple} as the EPR-Bell scenario \cite{epr,Bell1964,CHSH}; it is the most fundamental setup which allows for nonlocality, defined as an inequivalence between the statistics of LHVM and NOSIG.

To investigate the distinctions between LHVM, QM, and NOSIG models we contrast the different limitations they imposed on possible gedankenexperimental statistics. Finite and complete sets of inequalities have been determined for both NOSIG and LHVM in many multipartite and multichotomic scenarios \cite{GisinQuadChatomic,GisinNChotomic,WolfMultipartiteBell,WolfPolytopes}, however identification of the restrictions due to QM remains an open problem, even when limited to only the EPR-Bell scenario. A necessary and sufficient characterization of QM has been identified for the {\em 4-Space} of correlations within the EPR-Bell scenario, in which all marginal expectation values, e.g. \(\langle A_i\rangle\), are taken to be identically zero. We refer to those conditions as the \({\mbox{TLM}}^{(4)}\) inequalities, after Tsirelson \cite{LTM_T}, Landau \cite{LTM_L}, and Masanes \cite{LTM_M}) who each derived an equivalent form of them independently. Efforts to address the full {\em 8-Space} of both marginal and joint probabilities have been less developed. Tsirelson shared a theorem on the matter in 1980 \cite{tsirelson80} followed by the hierarchy of semi-definite-programming tests of Navascu\'{e}s, Pironio, and Ac\'{i}n (NPA) in 2007 \cite{NPA07,NPA08}, which converge in their infinite limit to a complete characterization of QM. Application of the NPA algorithm to the first-level of the hierarchy yielded the \({\mbox{NPA}}^{(8)}\) inequalities, the first (and heretofore only) known inequalities in 8-Space. The \({\mbox{NPA}}^{(8)}\) inequalities exactly identify statistical behaviors consistent with the principle of Macroscopic Locality \cite{MacroscopicLocality}. Those conditions have been used since as a `gold standard' to approximately delineate \({\mbox{QM}}^{(8)}\), such as in Ref. \cite{recover}, but one should not mistake them for complete.  Here we have introduced our superscript notation to indicate the probability space acted on by a given set of inequalities \footnote{The Bell inequalities refer only to correlations, so we refer to them as \({\mbox{LHVM}}^{(4)}\), in contrast to \({\mbox{NOSIG}}^{(8)}\).}.

In this paper we obtain a new characterization of \({\mbox{QM}}^{(8)}\) distinct from, and more complete than, \({\mbox{NPA}}^{(8)}\). To do so we requisitioned Theorem 2 of Ref. \cite{tsirelson80}, which was published without proof \footnote{Ref. \cite{tsirelson80} makes references to proofs that would be published elsewhere. Regrettably, Tsirelson's efforts to publish a followup article were obstructed by growing institutional antisemitism in Soviet Russia.}. In the process we identified a critical mathematical error which had been heretofore unnoticed. In this paper we construct the algorithm from first principles anew, thus providing both proof and correction to Tsirelson's theorem. The specific repaired expressions appear in \eq{eq:correction} here. For pedagogical purposes, as well as to facilitate a comparison to \({\mbox{NPA}}^{(8)}\), we develop the algorithm primarily within the EPR-Bell bipartite dichotomic scenario. We stress, however, that this algorithm naturally generalizes to any binary setup and is always simulated by shared qubits.
\section{II. The Algorithm}

Consider the following highly-general question: Given an arbitrary 8-Space measure $\exv{Z}$, weighted by real parameters $c_i$ such that
\begin{align}
\label{eq:expec}
&Z  \equiv \;\; c_1 A_0 + c_2 A_1 + c_3 B_0 + c_4 B_1 +\\
&\nonumber c_5 \,A_0\cdot B_0 + c_6\, A_1\cdot B_0 + c_7 \,A_0\cdot B_1 + c_8 \,A_1\cdot B_1
\end{align}
what is the upper limit on $\exv{Z}$; what is $\exv{Z}_{\rm Max}=$? 

For instance, choosing \(1=c_5=c_6=c_7=-c_8\) with all the other \(c_i=0\) yields \eq{eq:simple}. Determining upper bounds for such linear combinations of expectation values is a relatively straightforward problem in both LHVM and NOSIG models, as the inequalities which perfectly define them in 8-Space are known \cite{scarinilecture,GisinQuadChatomic,GisinNChotomic,WolfMultipartiteBell,WolfPolytopes} \footnote{Maximizing a linear function of expectation values subject to given linear inequalities is known as Semi-Definite Programming (SDP).}. However, since no such perfectly-defining inequalities exist yet for QM, a unique algorithm is called for, which we present here.

First we analogize the scenario's measurement apparatuses into quantum measurement operators. The spatial separation requirement is enforced in the condition that the operators of the different parties \(A_i\), \(B_i\),... commute with each other. We choose to put the operators in separate Hilbert spaces \cite{connes,physicalTsirelsonProblem}. The $\pm $1 possible measurements are embedded in the operators' eigenvalues, e.g. spin projectors to the x and y spin axes. Tsirelson proved that a 2-dimensional Hilbert space for each party was sufficient to emulate any general binary correlation behavior \cite{tsirelson80}. Without loss of generality we can consider only the {\em difference} in angle between a party's projective measurements, such that we can impose a reflection symmetry for dichotomic scenarios \footnote{For multichotomic scenarios one could set \(A_0,B_0,C_0\ldots\) and \(A_1,B_1,C_1\ldots\) as per \eq{eq:setup}, and define \(A_{\ge 2},B_{\ge 2},C_{\ge 2}\ldots\) in terms of {\em two} angles each, corresponding to arbitrary positions on their local Bloch spheres.}, i.e. 
\begin{align}\label{eq:setup}\begin{split}
&A_0\equiv \left(\cos (\theta_A) \sigma_x+\sin (\theta_A) \sigma_y\right) \otimes \mathfrak{1}\otimes \mathfrak{1}\ldots\\
&A_1\equiv \left(\cos (\theta_A) \sigma_x-\sin (\theta_A) \sigma_y \right)\otimes \mathfrak{1}\otimes \mathfrak{1}\ldots\\
&B_0\equiv \mathfrak{1} \otimes \left(\cos (\theta_B) \sigma_x+\sin (\theta_B) \sigma_y\right)\otimes \mathfrak{1}\ldots\\
&B_1\equiv \mathfrak{1} \otimes \left(\cos (\theta_B) \sigma_x-\sin (\theta_B) \sigma_y\right)\otimes \mathfrak{1}\ldots
\end{split}\end{align}

The quantum bound is, equivalently, the largest possible measurement value of \(Z\), the supremum of the largest eigenvalue, the global maximum of the right-most root of the secular equation. For the 8-Space \(Z\) defined in \eq{eq:expec} the secular equation has four roots, each parametrized by only the two variables, $\theta_A$ and $\theta_B$. Writing the characteristic polynomial of \(Z\) in terms of the variable \(m\) we identify the quantum bound for \eq{eq:expec} as
\begin{align}\begin{split}\label{eq:intractable}
\exv{Z}_{\rm Max} = \sup(m) \quad{\rm where}\quad\exists\;\norm{u}=\norm{v}=1  \\
\text{such that\quad}m^4+\mu _2  m^2+\mu _3 m+\mu _4 \;=\; 0 
\end{split}\end{align}
Here we are using a {\em corrected} form of Tsirelson's compact notation \cite{tsirelson80} such that
\begin{align}\label{eq:mus}
 &\mu_2=-\left(|e|^2+|f|^2\right)-2\left(|g|^2+|h|^2\right)\,,\\
 &\nonumber\mu_3=4\, \re{f h \bar{g} + e \bar{h} \bar{g}}\,,\\
 &\nonumber\mu_4=|e|^2|f|^2+\left(|g|^2-|h|^2\right)^2-2\,\re{f \bar{e} h^2+e f \bar{g}^2}\,,
\end{align}
\begin{align}\begin{split}\label{eq:correction}
 &e=u v c_5+v \bar{u} c_6+u \bar{v} c_7+\bar{u} \bar{v} c_8\,, \\
 &f=u \bar{v} c_5+\bar{u} \bar{v} c_6+u v c_7+v \bar{u} c_8\,, \\
 &g=u c_1+\bar{u} c_2\,,\quad h=v c_3+\bar{v} c_4\,,
\end{split}\end{align}
where \(u\) \& \(v\) are complex unitary variables, such that
\begin{align}\begin{split}
 &u=\cos(\theta_A)+i\sin(\theta_A)\,,\\
 &v=\cos(\theta_B)+i\sin(\theta_B)\,. \\
\end{split}\end{align}
and the over-bar notation in \eq{eq:mus} and \eq{eq:correction} is understood to indicate complex conjugation.

\begin{table*}[hbtp]
\caption{\label{tab:bounds}bounds (Number-Valued and Function-Valued): \({\mbox{TB}}^{(4)}\) is the well-known Tsirelson's bound, and \({\mbox{QB}_1}^{(4)}\) is its function-valued generalization. \({\mbox{QB}_3}^{(8)}\) surprisingly coincides with \(\text{LHVM}_\text{Max}<\text{NOSIG}_\text{Max}\) for all \(1\leq |x|\leq 2\). \({\mbox{QB}_3}^{(8)}\) is the most powerful of the bounds, alone capable of excluding by almost twice the volume excluded by all other known bounds combined, including \({\mbox{NPA}}^{(8)}\).}
\begin{ruledtabular}
\begin{tabular}{lrrrrrrrrccc}
 Name & $c_1$ & $c_2$ & $c_3$ & $c_4$ & $c_5$ & $c_6$ & $c_7$ & $c_8$ & \(\text{LHVM}_\text{Max}\) & \(\text{QM}_\text{Max}\) & \(\text{NOSIG}_\text{Max}\) \\
\colrule 
\({\mbox{TB}}^{(4)}\) & 0 & 0 & 0 & 0 & 1 & 1 & 1 & -1 & \({2}\) & \({2\sqrt{2}}\) & \({4}\) \\
\({\mbox{QB}_1}^{(4)}\) &  0 & 0 & 0 & 0 & 1 & 1 & 1 & x & \(|x+1|+2\) & \({\begin{cases}
  x+3 & \forall\; x\geq -\frac{1}{3} \\
  \sqrt{\frac{x^3-3 x^2+3 x-1}{x}} & \forall\; x\leq -\frac{1}{3}
 \end{cases}}\) & \(|x|+3\) \\
 \colrule 
 & 1 & 0 & 0 & 0 & 1 & 1 & 1 & -1 & 3 & \(\sqrt{10}\) & 4 \\
 \({\mbox{QB}_2}^{(8)}\) & x & 0 & 0 & 0 & 1 & 1 & 1 & -1 & \(|x|+2\) & \({\begin{cases}
  |x|+2 & \forall\; |x|\geq 2 \\
  \sqrt{2x^2+8} & \forall\; |x|\leq 2
 \end{cases}}\) & \(\begin{cases}
  |x|+2 & \forall\; |x|\geq 2 \\
  4 & \forall\; |x|\leq 2
 \end{cases}\)\\
 \colrule 
 & 1 & 1 & -1 & 0 & 1 & 1 & 1 & -1 & 3 & \(3\) & 4 \\
 \({\mbox{QB}_3}^{(8)}\) & x & x & -x & 0 & 1 & 1 & 1 & -1 & \(\begin{cases}
  3|x|-2 & \text{for } |x|\geq 2 \\
  |x|+2 & \text{for } |x|\leq 2
 \end{cases}\) & \({\begin{cases}
  3|x|-2 & \forall\; |x|\geq 2 \\
  |x|+2 & \forall\; 1\leq |x|\leq 2 \\
  \frac{x^2}{x^2-1}+\sqrt{\frac{3 x^4-10 x^2+8}{\left(x^2-1\right)^2}} & \forall\; |x|\leq 1
 \end{cases}}\) & \(\begin{cases}
  3|x|-2 & \forall\; |x|\geq 2 \\
  4 & \forall\; |x|\leq 2
 \end{cases}\)
\end{tabular}
\end{ruledtabular}
\end{table*}

\eq{eq:intractable} requires the analytically-intractable steps of identifying the largest solution to a quartic equation as well as maximizing over two variables simultaneously \footnote{The maximization would be over more than two variables if one is considering more than two parties or more than two measurement choices.}. We bypass both obstacles through the use of {\em for all} statements, shifting the variation from \(u\) and \(v\) to only \(m\). Observe that the characteristic polynomial has a definite positive leading coefficient (namely 1), and thus our operator's largest eigenvalue corresponds to that root of the polynomial at which all the derivatives are non-negative. As such, the quantum bound is given by the alternative formulation
\begin{align}
\nonumber\exv{Z}&_{\rm Max} = \inf(m)\quad\text{such that}\quad\forall\; \norm{u}=\norm{v}=1\text{ :}\\
&\begin{array}{lr}
  &m^4+\mu _2 m^2+\mu _3  m+\mu _4\,\geq\, 0  \\
  {\rm and}&\; 4 m^3+2 \mu _2  m+\mu _3\,\geq\, 0\\
  {\rm and}&\; 6 m^2+\mu _2 \,\geq\, 0\\
  {\rm and}&\; m\,\geq\, 0 
\end{array}
\end{align}
which is now suitable for analytic analysis. Note that the number of positivity constraints scales like two to the number of parties, and does not vary with multichotomic generalization.
\section{III. New quantum bounds}

Our ambition was to study those sets of \(c_i\) where the LHVM and NOSIG bounds do not coincide. We decided to examine all possible unitary or zero values for the eight \(c_i=-1,0,1\). A naive approach would be the computation and collection of all $3^8=6561$ sets, but through various physical symmetries (see Appendix B) we reduced those to $98$ inequivalent candidates, of which only three yield a gap between LHVM and NOSIG, listed in \tab{tab:bounds}. \({\mbox{TB}}^{(4)}\) there corresponds to \eq{eq:simple}, but other two number-valued bounds are new to the 8-Space analysis.

The algorithm's most powerful feature is not its yielding  of number-valued quantum bounds; other numerical algorithms can do so more efficiently \footnote{As an example of an efficient numerical algorithm one could compute the expectation value of the ``generic'' measurement operators above applied to some some ``generic'' bipartite state. Numerical maximization would be performed over the measurement angles as well as the amplitudes and phases of the generic state.}. The more significant reward of this {\em analytical} algorithm is that, when paired with the symbolic algebra prowess of Mathematica\texttrademark, our algorithm unleashes a new class of {\em function-valued} quantum bounds, such as \({\mbox{QB}_1}^{(4)}\),\({\mbox{QB}_2}^{(8)}\), and \({\mbox{QB}_3}^{(8)}\) in \tab{tab:bounds}.

Interestingly \({\mbox{QB}_3}^{(8)}\) demonstrates zero non-locality beyond the LHVM models for \(|x| > 1\) despite the overhead permitted by \(\text{NOSIG}_\text{Max}\) for \(|x| < 2\). This would allow for an ``inverse Hardy''-type \cite{hardy,hardyinfocaus} all-or-nothing test of Quantum Mechanics, in that the presence of any non-locality would not vindicate Quantum Mechanics, but rather contradict it for the parameter region in question.
\section{IV. Volume Analysis}

We can now contrast the function-valued quantum bounds in \tab{tab:bounds} with the known quantum bounds \({\mbox{TLM}}^{(4)}\) and \({\mbox{NPA}}^{(8)}\). Both \({\mbox{TLM}}^{(4)}\) and \({\mbox{NPA}}^{(8)}\) have the same form, namely that for all \(i\) and \(j\),
\begin{align}
\left|\mathfrak{f}_{(0,0)}+\mathfrak{f}_{(0,1)}+\mathfrak{f}_{(1,0)}+\mathfrak{f}_{(1,1)}-2\mathfrak{f}_{(j,k)}\right|\leq \pi
\end{align}
where for \({\mbox{TLM}}^{(4)}\), as derived by Masanes \cite{LTM_M}
\begin{align}
\mathfrak{f}_{(j,k})=\arcsin{\left\langle A_j B_k\right\rangle}
\end{align}
and for \({\mbox{NPA}}^{(8)}\) the authors found \cite{NPA07,NPA08}
\begin{align}
\mathfrak{f}_{(j,k})=\arcsin{\left(\frac{\left\langle A_j B_k\right\rangle -\left\langle A_j\right\rangle  \left\langle B_j\right\rangle
}{\sqrt{\left(1-\left\langle A_j\right\rangle {}^2\right) \left(1-\left\langle B_k\right\rangle {}^2\right)}}\right)}.
\end{align}
Recall that \({\mbox{TLM}}^{(4)}\) is complete in 4-Space, whereas \({\mbox{NPA}}^{(8)}\) is incomplete in 8-Space. Note that by setting the marginal expectations to zero in \({\mbox{NPA}}^{(8)}\) one recovers \({\mbox{TLM}}^{(4)}\), and therefore \({\mbox{TLM}}^{(4)}\) is wholly included within \({\mbox{NPA}}^{(8)}\).

We quantify the tightness of a set of criteria using a technique called {\em Volume Analysis}, first introduced in Ref. \cite{cabello05}; see therein for physical interpretations of the volume. As an informal estimate of the completeness of function-valued bounds in general we contrast the volume of \({\mbox{QB}_1}^{(4)}\) to both the weaker \({\mbox{TB}}^{(4)}\) and the stronger, complete, \({\mbox{TLM}}^{(4)}\) in 4-Space. The 4-Space Volume Analysis for \({\mbox{TB}}^{(4)}\) and \({\mbox{TLM}}^{(4)}\) has already been done in Ref. \cite{cabello05}, however the analytic value of the volume for \({\mbox{TB}}^{(4)}\) is original. We find that the function-valued \({\mbox{QB}_1}^{(4)}\) results in a significant tightening compared to the number-valued \({\mbox{TB}}^{(4)}\), in that \({\mbox{QB}_1}^{(4)}\) more than halves the excess volume of \({\mbox{TB}}^{(4)}\) relative to \({\mbox{TLM}}^{(4)}\).

\begin{table}[ht]
\caption{\label{tab:4space}Results of 4-Space Volume Analysis. The entirety of 4-Space has volume \(2^4=16\), and the tabulated values are understood as multiples of \({2^4}\). Note that the all points in 4-Space fit within \({\mbox{NOSIG}}^{(8)}\).}
\begin{ruledtabular}
\begin{tabular}{llll}
\({\mbox{TB}}^{(4)}\) & \({\mbox{QB}_1}^{(4)}\) & \({\mbox{TLM}}^{(4)}\) & \({\mbox{LHVM}}^{(4)}\)  \\
\(\frac{48\sqrt{2}-65}{3}\approx 0.961\) &\(\approx 0.938\) & \(\frac{3\pi^2}{32}\approx 0.925\) &  \(\frac{2}{3}\approx 0.667\)
\end{tabular}
\end{ruledtabular}
\end{table}

To analyze a function-valued bound such as \({\mbox{QB}_1}^{(4)}\) we compose a Boolean function of \(x\) that takes on the value 1 iff the inequality is violated, and returns 0 otherwise. A {\em point}, i.e. a set of hypothetical statistics, is within the bound iff the integral over \(x\) from \(-\infty\) to \(\infty\) is precisely 0. We used a Monte Carlo method to estimate the volume, sampling many random points to determine the approximate fraction contained within the criteria. 

To construct an approximation for \({\mbox{QM}}^{(8)}\) we combined all available quantum bounds, namely \({\mbox{NPA}}^{(8)}\), \({\mbox{QB}_2}^{(8)}\), and \({\mbox{QB}_3}^{(8)}\) and applied them to random nonlocal points, i.e. statistics rejected by \({\mbox{LHVM}}^{(4)}\) but permitted by \({\mbox{NOSIG}}^{(8)}\). We do not include \({\mbox{QB}_1}^{(4)}\) because its restrictions are wholly contained within \({\mbox{TLM}}^{(4)}\) which in turn is contained entirely within \({\mbox{NPA}}^{(8)}\). Using this model we found an upper bound for the true quantum volume in 8-Space of \(1083.8\times\frac{2^8}{8!}\). We also originally and analytically determined the volumes of \({\mbox{LHVM}}^{(4)}\) and \({\mbox{NOSIG}}^{(8)}\) within the \(2^8=256\) total volume of 8-Space. 

\begin{table}[b]
\caption{\label{tab:8space}Results of 8-Space Volume Analysis. The entire probability space, within which No-Signaling theories are contained, has total volume \(2^8=256\), and the tabulated values are given as multiples of \({2^8}/{8!}\).  The volume associated with \({\mbox{QM}}^{(8)}\) below is necessarily an overestimate, as the composite model used for \({\mbox{QM}}^{(8)}\) is only a collection of incomplete criteria.}
\begin{ruledtabular}
\begin{tabular}{cccc}
\({\mbox{NOSIG}}^{(8)}\) & \({\mbox{NPA}}^{(8)}\) & \({\approx\mbox{QM}}^{(8)}\) & \({\mbox{LHVM}}^{(4)}\)  \\
\(=1088\) & \(\approx 1086\) & \(\lesssim 1084\) & \(=1024\)
\end{tabular}
\end{ruledtabular}
\end{table}

We found \({\mbox{QB}_3}^{(8)}\) to be the most powerful bound, in that it alone has a volume of \(\approx 1084\times\frac{2^8}{8!}\), apparently singularly excluding 99\% of the nonlocal points excluded by the best composite model. Further composite models without \({\mbox{QB}_3}^{(8)}\) all had volumes larger than \(1085\times\frac{2^8}{8!}\). We also found the volume of \({\mbox{NPA}}^{(8)}\) to be \(\approx 1085.8\times\frac{2^8}{8!}\), significantly less than the volume of either \({\mbox{QB}_2}^{(8)}\) or \({\mbox{QB}_3}^{(8)}\) alone. We did identify some points that were forbidden by \({\mbox{NPA}}^{(8)}\) but not by \({\mbox{QB}_2}^{(8)}\) nor \({\mbox{QB}_3}^{(8)}\), however these exceptional statistics comprise a relatively miniscule volume, numerically capped by at most \(0.1\times \frac{2^8}{8!}\)\,. To summarize: \({\mbox{QB}_3}^{(8)}\) is significantly more restrictive than \({\mbox{NPA}}^{(8)}\), and \({\mbox{NPA}}^{(8)}\) is almost entirely contained within \({\mbox{QB}_3}^{(8)}\). 

It would be gratifying to have some lower bound for the true volume of \({\mbox{QM}}^{(8)}\). Towards this goal we constructed a pseudo-quantum polytope defined by the 16 extremal points of \({\mbox{LHVM}}^{(4)}\) together with the quantum analog of the 8 nonlocal extremal points of \({\mbox{NOSIG}}^{(8)}\). By a variation of the convex hull method we determined that the volume of this pseudo-quantum polytope is somewhat larger than \(\approx 1036\times\frac{2^8}{8!}\). This lower bound for \({\mbox{QM}}^{(8)}\) is so far below the upper bound of \(1083.8\times\frac{2^8}{8!}\) such that it does not offer much insight. Polytope construction with additional extremal points would undoubtedly yield a better lower bound, however such higher-precision efforts were effectively unfeasible using our non-professional computational methods. 

\section{V. Conclusion}

Thus we have rederived, repaired, and repurposed Tsirelson's theorem \cite{tsirelson80} into an algorithm for obtaining analytic quantum bounds. In our formulation it is explicitly suitable even for the study of multipartite nonlocality, and in principle extendable also to multichotomic scenarios, although we restricted our focus in this paper to the 8-Space of the bipartite dichotomic EPR-Bell scenario. We identified new quantum bounds relevant to this highly-studied scenario, and then used those new bounds to perform an approximate volume analysis for \({\mbox{QM}}^{(8)}\).

This algorithm is only one path towards approaching a genuinely complete characterization, the convergent hierarchy of Navascu\'{e}s et al \cite{NPA08} is another. Still missing, but eagerly anticipated, is the discovery of complete inequalities defining \({\mbox{QM}}^{(8)}\), akin to the discovery of \({\mbox{TLM}}^{(4)}\) for \({\mbox{QM}}^{(4)}\). 

We thank B.S. Tsirelson of Tel Aviv University for his input and discussions that enhanced this paper considerably, as well as Roger Colbeck of the Perimeter Institute for Theoretical Physics. We wish to thank the NSF for funding.

\bibliographystyle{apsrev4-1}
\bibliography{wolfePRAv1}

\onecolumngrid
\appendix \section{APPENDICES}
\subsection{APPENDIX A: The Algorithm as Computer Code}

Here we present the algorithm as above in a form of psuedo-code, generalized for mutlipartite scenarios restricted to binary output and dichotomic measurements, in the manner of \eq{eq:setup}.

\begin{enumerate}
	
\item  Accept input in the form of a linear function, for example:
\begin{align}\label{eq:input}
input=x(A_0\otimes \mathbf{1})+(A_0\otimes B_0)+(A_1\otimes
B_0)+(A_0\otimes B_1)-(A_1\otimes B_1)
\end{align}

\item Interpret statistical measures as a quantum operators, such that \(\otimes\) can be interpreted as the matrix Kronecker Product.
\begin{align}
Z=input:{\text{Party}[k]}_{\text{Apparatus}[n]}\to \cos (\theta_k) \sigma_x+\left(-1\right)^n \sin (\theta_k) \sigma_y
\end{align}

\item Obtain the characteristic polynomial of the operator \(Z\) and all of its non-constant derivatives.
\begin{align}
D_0&=\text{CharacteristicPolynomial}\left(Z\right) \text{ with respect to } m\\\nonumber
&= m^{(2^K)}+\mu_2\, m^{(2^K-1)}+\mu_3\, m^{(2^K-2)}\;...\;+\mu_{(2^K-1)}\, m+\mu_{(2^K)}\\\nonumber
D_{i+1}&\equiv\frac{\partial {D_i}}{\partial m}\\\nonumber
D_{final}&=D_{(2^K-1)}=(2^K-1)!\;m
\end{align}

\item Subsequent to algebraic simplification we rewrite the all expressions in terms cosine exclusively. This is preliminary to the next step where we use the cosine as a proxy for angle since we are only considering \(0\le\theta\le\pi\).
\def\col{\mathop{\text{Collect}}\limits} 
\begin{align}
Z&=Z:\sin{\theta_k}\to\sqrt{1-\cos^2{\theta_k}}
\end{align}

\def\inand{\mathop{\land}\limits}
\item The complete \(Conditions\) are formed by the logical union, ie AND, of all \(\{i:0\; .. \; 2^K-1\}\) positivity conditions.
\begin{align}
Conditions=\cup_{i}\left(\forall_{\cos{\theta_{k,n}}}{D_i\ge 0}\right)=\forall_{\cos{\theta_{k,n}}}\cup_{i}\left(D_i\ge 0\right)
\end{align}

\item The Quantum Bound is finally obtained by minimizing m subject to all \(Conditions\):
\begin{align}
QB=\inf_{m \in \mathbb{R} }{\left(m,\text{ such that }Conditions\right)}\\\nonumber
\end{align}

\end{enumerate}

\begin{samepage}

\subsubsection{Example Intermediate and Final Results from the Algorithm}
Should the reader wish to practice the algorithm as expressed above we reproduce below the positivity conditions and final Quantum Bound for the example input of \eq{eq:input}. They are as follows: for all values of \(\cos(\theta_A)\) and for all values of \(\cos(\theta_B)\), interpreting \(\sin{\theta_k}\to\sqrt{1-\cos^2{\theta_k}}\)\;the characteristic polynomial variable \( m\) must satisfy
\nopagebreak
\begin{align*}
 & m^4-
m^2 \left(2x^2+8\right)+(x^2-4)^2+64\left(\cos^2{\theta_A}\right)\left(\sin^2{\theta_A}\right)\left(x^2-4\cos^2{\theta_B}\right)\left(\sin^2{\theta_B}\right) \ge 0 \\
 &m^3-m \left(x^2+4\right)\ge 0 \\
 &3 m^2-(x^2+4)\ge 0 \\
 &m\ge 0
\end{align*}
One can readily demonstrate that the infimum of \(m\) satisfying all the above conditions is
\begin{align*}
	\begin{cases}
  |x|+2 & \text{for } |x|\geq 2 \\
  \sqrt{2x^2+8} & \text{for } |x|\leq 2
 \end{cases}
\end{align*}
as per \({\mbox{QB}_2}^{(8)}\) in Table I in the main text.
\end{samepage}

\subsection{APPENDIX B: Physical Symmetries}

In the bipartite and dichotomic EPR-Bell scenario \cite{epr,Bell1964,CHSH} there are exactly seven independent physically motivated symmetries. The first symmetry is the relabeling of the parties, Alice and Bob. The next two come from the permuting the indices of each party's measurement apparatuses. The remaining four come from permuting the binary outputs any of the measurements, a redefinition of success and failure in a sense. For statistics defined the eight parameters of 
\eq{eq:expec} the fundamental symmetries are given below.

For \(K\) parties, each having access to \(N\) choices of experimental apparatus, each experiment yielding one of \(D\) possible outcomes, the physical symmetries are as follows: There are \(K!\) permutations of party labeling-scheme, \(N!\) permutations of the apparatus labeling scheme for each of the \(K\) parties, and \(D!\) output labeling-schemes for each of the \(K\times N\) apparatus settings. The total number of equivalencies, therefore, is \(K!\;{N!}^K\;{D!}^{K N}\)\;. The total number of fundamental symmetries is composed of the identity, \(K!-1\) party permutations, \(K\left(N!-1\right)\) apparatus permutations, and \(K N \left(D!-1\right)\) output permutations.

\begin{table}[hb]
\caption{\label{tab:sym} The 7 non-identity symmetries pertinent to EPR-Bell scenario. These symmetries form a group with \(2^7=128\) equivalencies.}
\begin{ruledtabular}
\begin{tabular}{llll}
Permutation of the parties : \(K!-1\) & \(\begin{cases}
	A_i\leftrightarrows B_i &\quad\;\qquad\begin{array}{lll} C_1\leftrightarrow C_3 &\quad C_2\leftrightarrow C_4 &\quad C_6\leftrightarrow C_7 \end{array}\end{cases}\) & \qquad \\\colrule
 Permutations of the apparatuses : \(K\left(N!-1\right)\) &\(\begin{cases} A_0\leftrightarrows A_1 &\quad\qquad\begin{array}{ccc} C_1\leftrightarrow C_2 &\quad C_5\leftrightarrow C_6 &\quad C_7\leftrightarrow C_8 \end{array}\\ 
 B_0\leftrightarrows B_1 &\quad\qquad\begin{array}{ccc} C_3\leftrightarrow C_4 &\quad C_5\leftrightarrow C_7 &\quad C_6\leftrightarrow C_8 \end{array}\end{cases}\)\\\colrule
 Permutations of the output : \(K N \left(D!-1\right)\) &\(\begin{cases}A_0\leftrightarrows -A_0 &\,\qquad\begin{array}{ccc} C_1\leftrightarrow -C_1 & \;C_5\leftrightarrow -C_5 &\; C_7\leftrightarrow -C_7 \end{array}\\
 A_1\leftrightarrows -A_1 &\,\qquad\begin{array}{ccc} C_2\leftrightarrow -C_2 &\; C_6\leftrightarrow -C_6 & \;C_8\leftrightarrow -C_8 \end{array}\\
 B_0\leftrightarrows -B_0 &\,\qquad\begin{array}{ccc} C_3\leftrightarrow -C_3 &\; C_5\leftrightarrow -C_5 &\; C_6\leftrightarrow -C_6 \end{array}\\
 B_1\leftrightarrows -B_1 &\,\qquad\begin{array}{ccc} C_4\leftrightarrow -C_4 &\; C_7\leftrightarrow -C_7 &\; C_8\leftrightarrow -C_8 \end{array}
 \end{cases}\)
\end{tabular}
\end{ruledtabular}
\end{table}

\end{document}